\documentclass[12pt]{article}
\usepackage{figlatex}
\usepackage{graphics}
\usepackage{scicite}

\usepackage{times}

\newenvironment{sciabstract}{%
\begin{quote} \bf}
{\end{quote}}

\newcounter{lastnote}
\newenvironment{scilastnote}{%
\setcounter{lastnote}{\value{enumiv}}%
\addtocounter{lastnote}{+1}%
\begin{list}%
{\arabic{lastnote}.}
{\setlength{\leftmargin}{.22in}}
{\setlength{\labelsep}{.5em}}}
{\end{list}}

\title{ 
Detection of a Large Scale Structure of Intracluster Globular Clusters in the Virgo Cluster } 

\author
{Myung Gyoon LEE,$^{1\ast}$ Hong Soo PARK,$^1$ Ho Seong HWANG$^{1,2}$\\
\\
\normalsize{$^{1}$Astronomy Program, Department of Physics and Astronomy,}\\
\normalsize{Seoul National University, Seoul 151-742, Korea}\\
\normalsize{$^{2}$  CEA Saclay/Service d'Astrophysique, F-91191 Gif-sur-Yvette, France}\\
\\
\normalsize{$^\ast$To whom correspondence should be addressed; E-mail:  mglee@astro.snu.ac.kr.}
}

\date{}

\begin{document} 

\baselineskip24pt

\maketitle 
One-sentence summary:
We find a large scale structure of intracluster globular clusters, using the map for the globular clusters in the Virgo cluster of galaxies.


\begin{sciabstract} 
Globular clusters are found  usually in galaxies and they
are an excellent tracer of dark matter.
Long ago it was suggested that there may exist intracluster globular clusters (IGCs)
bound to a galaxy cluster rather than to any single galaxy. 
Here we present a map showing the large  scale distribution of globular clusters over the entire Virgo cluster. 
It shows 
that IGCs are found out to 5 million light years from the Virgo center, and that they are concentrated in several substructures much larger than galaxies.
These objects might have been mostly stripped off  from low-mass dwarf galaxies.  
\end{sciabstract}



Almost six decades ago it was suggested that there may exist stars and interstellar medium 
  between galaxies in clusters of galaxies\cite{zwi51,van56}. 
To date there is accumulating evidence for the existence of 
  diffuse optical stellar light, planetary nebulae, resolved red giant stars,   and diffuse hot 
  X-ray emitting  gas   between galaxies 
  in nearby galaxy clusters\cite{boh94,dur02,fel04,mih05,cas09}. 
Some of these objects are not bound to any single galaxy, 
 but are gravitationally controlled by the potential of the galaxy cluster itself. 
These are called intracluster objects. 

Later it was suggested that globular clusters should be stripped off from galaxies 
  in a galaxy cluster and that there should be a cluster-wide population of intracluster globular clusters (IGCs) in galaxy clusters\cite{whi87,muz87,wes95}. 
However, so far there has been no indication for IGCs in the Coma cluster,  
and only a small number of IGCs were found in similar studies of other galaxy clusters 
because of their relatively 
shallow photometric limits or small area coverage\cite{jor03,mar03,tam06,wil07a}. 
It is not yet known whether 
this  cluster-wide population of IGCs exists in any galaxy cluster or not. 

The Virgo cluster is the best target  to search for a cluster-wide population of IGCs, 
because it is the nearest massive galaxy cluster.
However, 
because of its largest angular extent (over ten degrees in the sky)\cite{bin85} 
and the faintness of its globular clusters 
it has been difficult to find and study IGCs 
over the entire Virgo cluster.

We present the results of a search for the globular clusters
  over the entire Virgo cluster using data from the Sloan Digital Sky Survey 
(SDSS)\cite{yor00,note1}.
The SDSS data are deep enough to study the bright globular
clusters in Virgo and
its survey area is wide enough to cover the entire Virgo cluster.

The Virgo cluster is located at a distance of 54 million light years (16.5 mega parsec); 
one degree in the sky corresponds to 940,000 light years (288 kilo parsec) at this distance.
The globular clusters at the distance of Virgo appear as point sources  in the SDSS images
thus they cannot be distinguished from faint foreground stars in our Galaxy in the images.
To overcome this problem and create a map of the globular clusters in Virgo, 
we used the photometry of the point sources in the SDSS Sixth Data Release\cite{ade08}. 

 We selected bright globular cluster candidates 
in a circular field with radius of 9 deg including the Virgo cluster,
 using the criteria for color and magnitude (reddening corrected):  $0.6< (g-i)_0 <1.3$ and $19.5< i_0 <21.7$ mag, 
 as marked by the box in Fig. 1.
This color range is similar to that used for selecting globular clusters in M87\cite{har09},
and corresponds to a metallicity range of
$-2.0 ^{<}_{\sim}  [{\rm Fe/H}]  ^{<}_{\sim} +0.4$\cite{lee08,note2}. 
The lower magnitude limit, $i_0<21.7$ mag 
is $\sim 1.5$ mag brighter than the peak magnitude in the luminosity function of globular clusters\cite{har09}. 
By using globular clusters brighter than $i_0 =21.7$ mag, 
we covered $\sim 13$ \% of the entire globular cluster population,
assuming that their luminosity function is represented by a Gaussian function with peak $i_0 =23.17$ mag\cite{har09}. 

We created a surface number density map for the globular clusters in Virgo 
after subtracting the contribution of foreground stars 
[including the Virgo overdensity \cite{jur08}] 
using the color-magnitude diagram method 
(details are described in the supporting online material) (Fig. 2A). 
The surface number density map for the globular clusters reveals
the existence of a diffuse large scale distribution of globular clusters (in cyan and green colors), 
  extending out to about 6 degrees (5 million light years) from the Virgo center,
   close to the boundary of the Virgo cluster.

 It also shows the presence of  several strong concentrations of globular clusters. 
Their locations correspond to the positions of bright elliptical galaxies 
  such as M87, M49, M60, M86/M84, and NGC 4636. 
However, the sizes of these substructures are much larger than individual galaxies. 
This population must include not only globular clusters in galaxies but also IGCs.

The large-scale distributions of globular clusters surrounding M87, M49, and M60 
  are similar in general to that of large scale X-ray emission seen in the ROSAT X-ray map (Fig. 2B). 
  However,   one diffuse component of globular clusters 
  in the west of M87 is not seen in the X-ray map, 
  while one diffuse X-ray component in the north of M87 is not found in the map of globular clusters.

There are tidal features around some bright galaxies: 
an elongated structure in the outer part of M87, and  bridge-like features
 between M87 and M86 (also between M87 and M89).
These features are  similar to the distribution of diffuse stellar light 
 found  in the deep optical image of the inner 1.5 deg$\times$1.5 deg region\cite{mih05}.

It is known that the color distribution for the globular clusters 
  in giant elliptical galaxies is bimodal, 
  showing that there are two populations: 
 blue and  red globular clusters 
(i.e., metal-poor and metal-rich globular clusters, respectively)\cite{lee03,pen06}. 
The large-scale distribution of the blue globular clusters in the Virgo cluster
 is much more extended than that of the red globular clusters (Fig. 3); 
  a similar trend  is seen around bright galaxies (most dramatically around M49 and M87).
Therefore the IGCs are probably  dominated by the blue globular clusters.
  
We derived the radial number density profiles for the globular clusters ($\sigma$ [number per square arcmin])
 as a function of radius 
  from the center of the Virgo cluster ($R$)\cite{note3} (Fig. 4). 
The radial number density profiles of the globular clusters show a change in the slope at $R \approx 40$ arcmin, 
  beyond which the slope gets  flatter.    
This break position is consistent with the value for the edge of the stellar halo of M87, $R=34$ arcmin, 
derived from the kinematic study of planetary nebulae in Virgo \cite{doh09}. 
Therefore the globular clusters at $R<40$ arcmin are considered to mostly belong to M87, 
while those at $R>40$ arcmin may be mostly IGCs. 
Double linear fits for all the globular clusters yield 
$\log(\sigma) = -1.49(\pm 0.09) \log R + 1.05(\pm0.03)$
for the outer region (40 arcmin $<R<6$ deg), and  
$-2.07(\pm 0.01) \log R + 1.99(\pm0.002)$
for the inner region ($R<40$ arcmin).
The slope for all the globular clusters at the outer region is similar to that for the dwarf elliptical galaxies, and is consistent with the lower limit predicted 
from numerical simulations, --1.5 to --2.5 \cite{bek06}.  
This result suggests that the IGCs may follow the dark matter distribution of the Virgo galaxy cluster. 
If the globular clusters in the core of Virgo undergo a transition 
from M87-dominated to Virgo-dominated, 
it is expected that their velocity dispersion profile for $R>40$ arcmin is relatively 
flat with a value that is a few hundred km s$^{-1}$ higher than that for the inner region\cite{cot01}.

The number density of the blue globular clusters is, on average, about twice as large as that of the red globular clusters for the range 40 arcmin $<R<$ 6 deg (Figs. 4B and 4C), indicating again
that the IGCs may be dominated by the blue globular clusters.
For the range 40 arcmin $< R < 6$ deg, we measured $N=1,500\pm300$  
IGCs with $19.5<i_0 <21.7$ mag.
This implies the total number of entire IGCs in that region 
is $N({\rm total}) \sim 11,900$, 
assuming that the luminosity function for IGCs is similar to that for globular clusters in galaxies. 

There are several models to explain the origin of globular clusters in giant elliptical galaxies,
but their origin is still  controversial \cite{lee10}.
Our finding that there is a wide distribution of IGCs and that  they are mostly blue (i.e., metal-poor) indicates that the major origin of these IGCs is low-mass dwarf galaxies\cite{mil07,bek08,coe09}.
This supports the mixture scenario for the origin of globular clusters in giant elliptical galaxies\cite{lee10}. According to this scenario, metal-poor globular clusters were formed mostly in low-mass dwarf galaxies and metal-rich globular clusters were formed later with stars in massive galaxies or in dissipational merging galaxies.
Elliptical galaxies grow via dissipational or dissipationless merging of galaxies and via accretion of many dwarf galaxies.
The IGCs in Virgo might have been stripped off from low-mass dwarf galaxies and are now
being accreted locally to nearby massive galaxies, and globally to the center of Virgo.



\smallskip


\begin{scilastnote}
\item  
We thank D. Geisler, M. Im and N. Hwang
for their careful reading the manuscript and useful comments,
 to the  anonymous referees for their comments that improved the original manuscript significantly.
This work was supported by the National Research Foundation of Korea (NRF) grant funded by
the Korea Government (MEST) (No. R01-2007-000-20336-0). 
Funding for the SDSS and SDSS-II has been provided by the Alfred P. Sloan Foundation, the Participating Institutions, the National Science Foundation, the U.S. Department of Energy, the National Aeronautics and Space Administration, the Japanese Monbukagakusho, the Max Planck Society, and the Higher Education Funding Council for England. The SDSS Web Site is http://www.sdss.org/. The SDSS is managed by the Astrophysical Research Consortium for the Participating Institutions. The Participating Institutions are the American Museum of Natural History, Astrophysical Institute Potsdam, University of Basel, University of Cambridge, Case Western Reserve University, University of Chicago, Drexel University, Fermilab, the Institute for Advanced Study, the Japan Participation Group, Johns Hopkins University, the Joint Institute for Nuclear Astrophysics, the Kavli Institute for Particle Astrophysics and Cosmology, the Korean Scientist Group, the Chinese Academy of Sciences (LAMOST), Los Alamos National Laboratory, the Max-Planck-Institute for Astronomy (MPIA), the Max-Planck-Institute for Astrophysics (MPA), New Mexico State University, Ohio State University, University of Pittsburgh, University of Portsmouth, Princeton University, the United States Naval Observatory, and the University of Washington.
\end{scilastnote}

\medskip

\noindent {\bf Supporting Online Material} 

\noindent www.sciencemag.org \\
SOM Text (Methods) \\
Figs. S1 and S2 \\ 
References


\clearpage


\begin{figure}
\includegraphics [width=\textwidth]{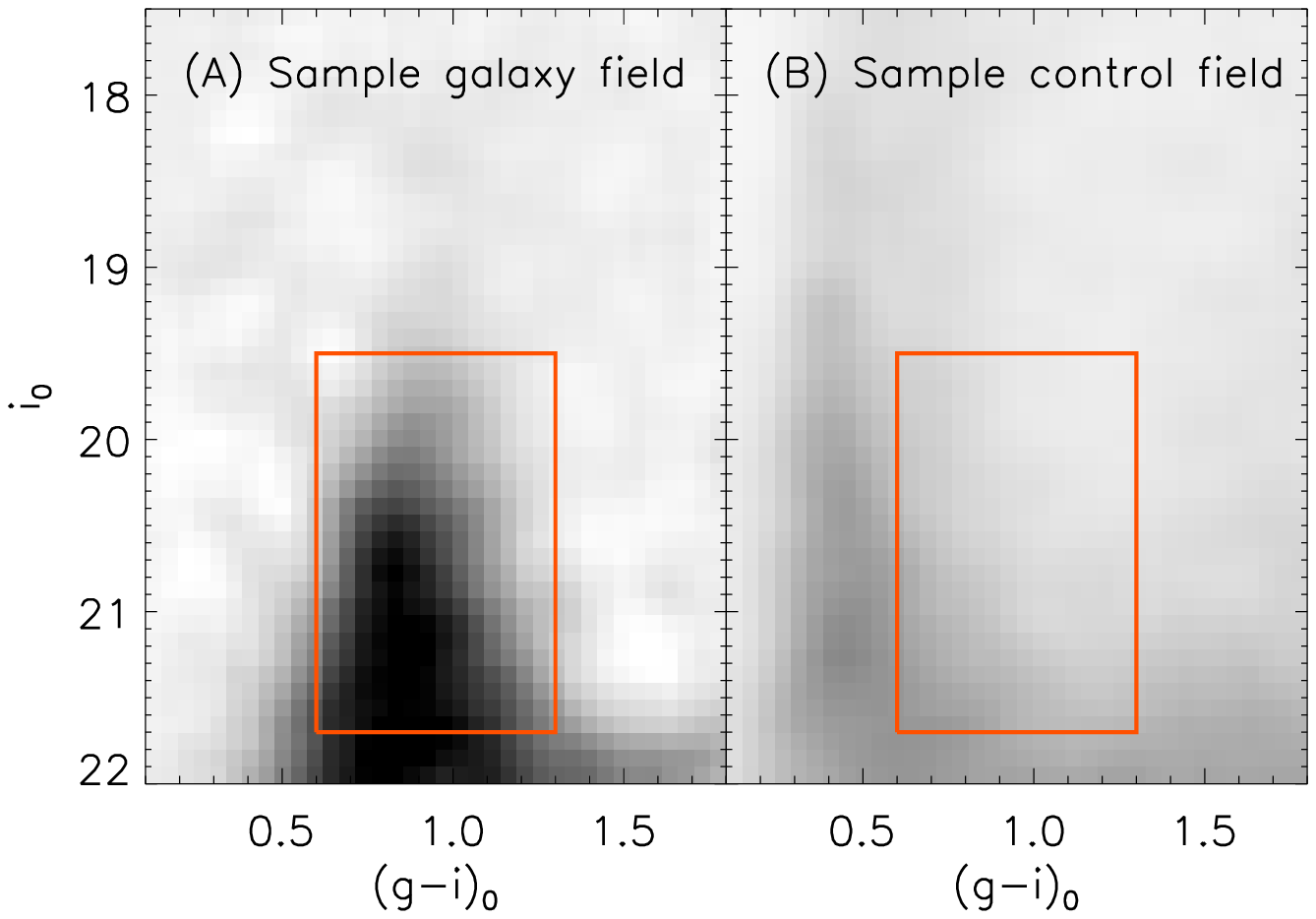} 
\caption{ 
$i_0 -(g-i)_0$ color-magnitude diagrams for the point sources in Virgo.
(A) A sample galaxy field composed of ten circular fields (radius 20 arcmin) centered on the brightest galaxies. 
The contribution of the foreground stars was subtracted. 
Most of the sources are globular clusters. 
(B) A sample control field located at 6 to 9 deg from the Virgo center, covering the same area as (A). Most of the sources are foreground stars belonging to our Galaxy.
Darker colors represent the higher number density.
The box represents the region used for selecting the bright globular cluster candidates in Virgo
in this study.
}
\label{fig1}
\end{figure}
\clearpage

\begin{figure}
\includegraphics [width=\textwidth]{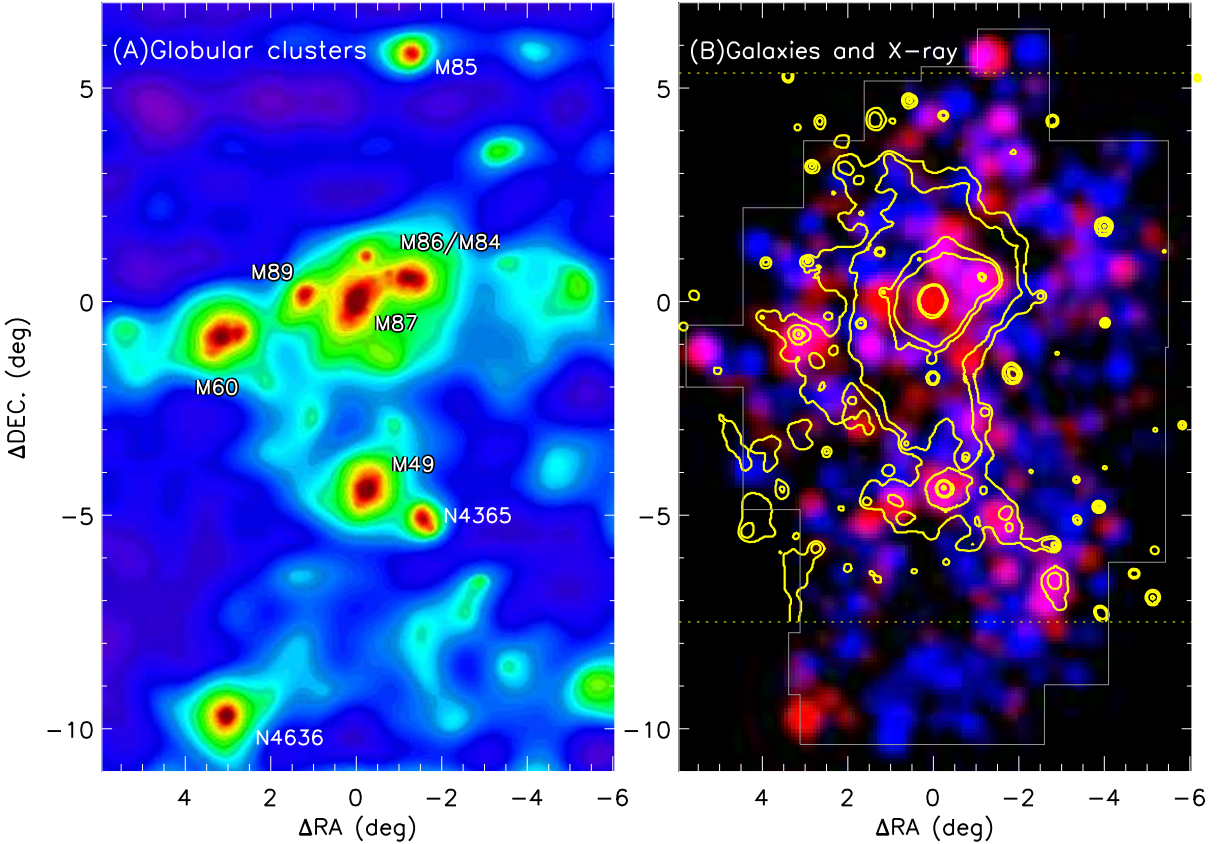} 
\caption{ 
Surface number density maps for the globular clusters in comparison with the spatial distribution of galaxies and  X-ray emission in the Virgo cluster region. 
(A) Globular clusters.
The levels are 0.012, 0.025, 0.06, and 0.12 per arcmin$^2$, respectively,
for cyan, green, yellow, and red boundaries.  
The positions of several bright elliptical galaxies are labeled.
(B) Galaxies and X-ray emission.
 The pseudo color map represents 
  a luminosity-weighted surface number density map of the early-type (red color) 
  and late-type galaxies (blue color)  that are members of Virgo. 
The yellow contours represent the contours of X-ray image for the Virgo cluster region  in ROSAT all-sky survey (in the hard energy band 0.4-2.4 keV)\cite{boh94}. North is up and east to the left.
}
\label{fig2}
\end{figure}
\clearpage

\begin{figure}
\includegraphics [width=\textwidth]{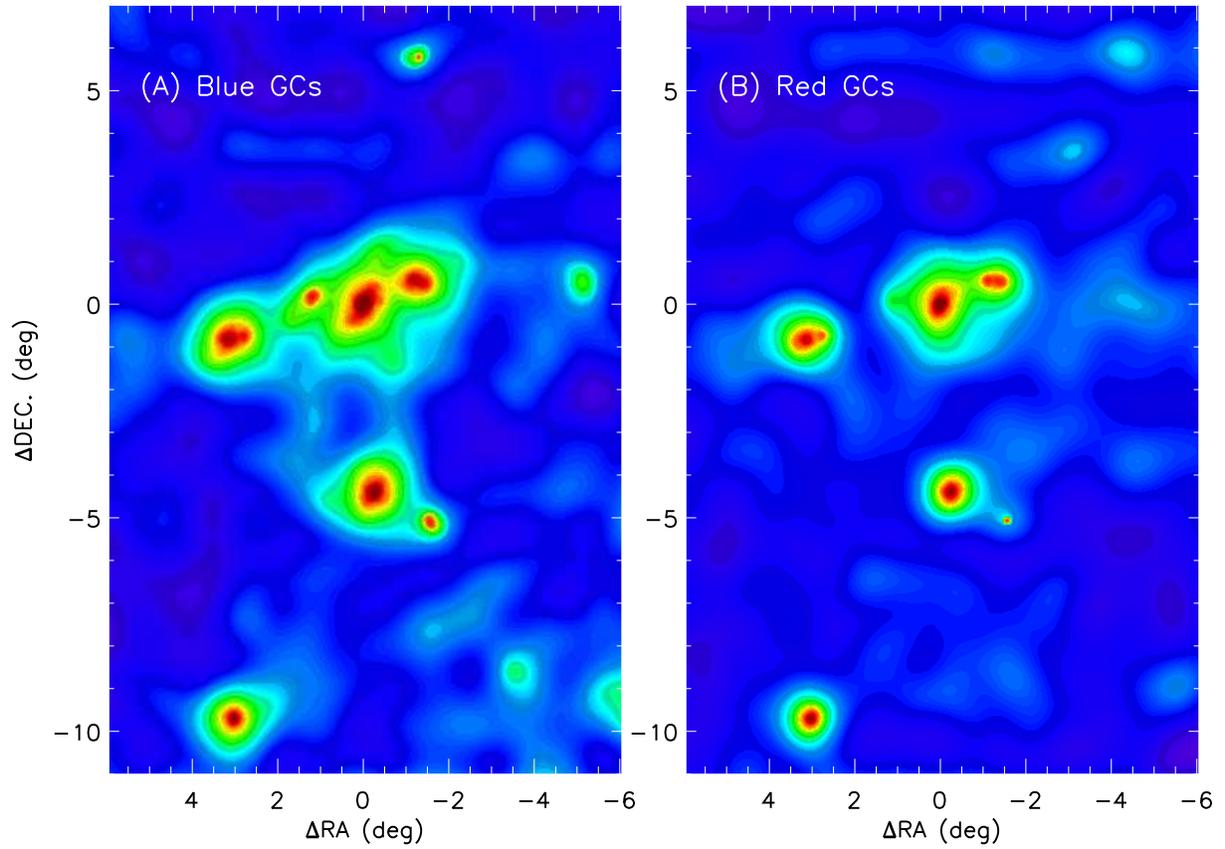} 
\caption{
Surface number density maps of the blue globular clusters ($0.6<(g-i)_0 \le 0.95$)(A) and
  the red globular clusters ($0.95 < (g-i)_0 < 1.3$)(B). The color for the boundary between the blue and red clusters, $(g-i)_0=0.95$,  corresponds to the metallicity 
  [Fe/H] $\approx -0.82$. The levels are the same as in Figure 2(A). 
}
\label{fig3}
\end{figure}
\clearpage

\begin{figure}
\includegraphics [width=\textwidth]{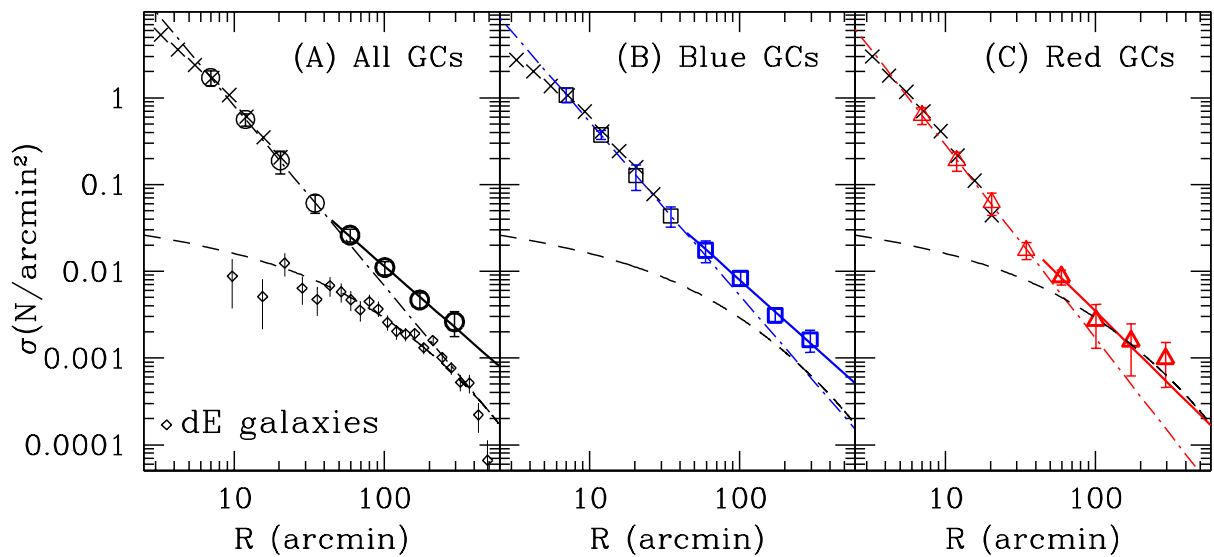} 
\caption{
Radial number density profiles for the globular clusters in Virgo,
obtained after masking out a circular region (radius $<5 R_{25}$) for each galaxy in the Virgo galaxy catalog
except for M87 to remove globular clusters in galaxies.
(A) All the globular clusters (circles). 
Dot-dashed lines and solid lines represent the linear fits for the inner region at $R<40$ arcmin, and the outer region at $R>40$ arcmin, respectively. 
Diamonds represent the radial number density profile 
  for the dwarf elliptical (dE) galaxies that are members of Virgo. 
Curved dashed lines represent a projected NFW profile\cite{nav97} for the Virgo mass distribution  given by McLaughlin\cite{mcl99}. 
(B) The blue globular clusters (squares), and (C) the red globular clusters (triangles).
Crosses represent the data including the fainter ($i<23.0$) globular clusters  at $R<30$ arcmin from M87 in given by Harris\cite{har09}, 
arbitrarily shifted to match the data for $10<R<20$ arcmin.
Note that they are in a good agreement with our data for the inner region.
}
\label{fig4}
\end{figure}
\clearpage


\noindent {\Large \bf SUPPORTING ONLINE MATERIAL: Detection of a Large Scale Structure of Intracluster Globular Clusters in the Virgo Cluster} 

\bigskip \noindent {\Large\bf Methods} \bigskip

Here we briefly describe the methods used to derive the surface
number density maps for globular clusters in the Virgo cluster, using the SDSS point source catalog.
We used the photometry of the point sources in the SDSS Sixth Data Release ({\it S1}) 
to select globular cluster candidates in  Virgo.

The SDSS catalog of point sources in the Virgo direction includes 
not only the Virgo globular clusters but also the objects that do not belong to the Virgo cluster.
There are two types of point sources that are not  members of Virgo, which are located
in the direction of Virgo:
1) foreground stars belonging to our Galaxy, and 2) unresolved background galaxies.  
Since we are dealing with relatively bright point sources, the contribution due to the latter is
negligible ({\it S2}). 
So we consider only foreground stars. 

We need to remove foreground stars and keep only Virgo globular clusters 
for making a surface number density map for globular clusters.
The foreground stars in Virgo are contributed by two types of objects. 
One is the disk and halo of our Galaxy, and the other is the Virgo overdensity 
discovered recently
in the SDSS ({\it S3}). 
 The Virgo overdensity is a large scale substructure covering over 1000 deg$^2$ of sky toward Virgo, and is estimated to be located at a distance of 6-20 kilo parsec. It 
may be a tidal stream or a low-surface brightness dwarf galaxy merging with the Milky Way ({\it S3}). 
Because of the existence of the Virgo overdensity overlapping with our field of interest, 
the conventional method of using the separate
control field at the similar galactic latitude does not work well for subtracting the contribution of these foreground stars. 
Therefore we adopted  a method of using the color magnitude
diagram (CMD) for foreground subtraction as follows.

We selected four fields in the sky to choose appropriate regions in the CMDs for 
selecting foreground stars 
and globular clusters in our analysis.
(a) The Virgo cluster field: a circular region with radius $R=6$ deg, centered on M87.
This field covers approximately the entire area of the Virgo cluster.
(b) The Control field: an annular region with radius $6<R<9$ deg, surrounding the Virgo
cluster field. This field represents a background field close to the Virgo cluster. 
It is used for subtracting the contribution of local foreground stars from the Virgo cluster field.
(c) Two Reference fields: One circular region (for the Virgo cluster field) and 
one annular region (for the Control field), both of which are centered at
(galactic longitude $l=80.32$ deg, galactic latitude $b=74.46$ deg).
They are at the same galactic latitude as the Virgo fields, but $\sim 200$ deg distant in galactic longitude from the Virgo fields. They have the same areas as (a) and (b), respectively. 
These reference fields are used for checking the contribution of general foreground stars in (a) and (b).

Then we constructed raw $i_0 -(g-i)_0 $ CMDs for the Virgo cluster field and its
reference field, and the Control field and its reference field,
as shown in Fig. S1(A), (B), (D), and (E). 
We subtracted the CMD for the corresponding reference field from the CMDs for the Virgo cluster field and the Control field, 
displaying the resulting CMDs in Fig. S1(C) and (F).

We used Fig. S1(C) and (F) for choosing the CMD regions  to make foreground star maps.
Fig. S1(C) includes the contribution of the foreground stars (the Virgo overdensity stars 
and the remaining disk and halo stars) and the globular clusters in Virgo, while
Fig. S1(F) includes the contribution of the foreground stars only.
These figures show that the stars in the Virgo overdensity are mostly bluer than the globular clusters.
The Virgo overdensity stars contaminate somewhat the bluer side of the blue globular cluster region, while it affects little the red globular cluster region.

We marked the regions for selecting foreground stars (as well as globular clusters) 
by boxes in Fig. S1. 
We chose a bright blue star region (BBS) with the same color range as the  blue globular cluster region,
and a faint blue star region (FBS) with the same magnitude range as the globular cluster region as in Fig. S1(C) and (F), 
and used the stars in these regions for creating a blue foreground map 
to be used for the blue globular clusters.
Then we chose a bright red star region (BRS) with the same color range as the red globular cluster region,
and a faint red region (FRS) with the similar magnitude range to that of the globular cluster region, 
and used the stars in these regions for creating a red foreground map to be used for the red globular clusters.   

We made the raw surface number density maps for globular cluster candidates (Fig. 2S(A), (D), and (G)) and foreground stars (Fig. 2S(B), (E), and (H)) using the objects inside the above regions in the CMDs.
The foreground map was made with a combination of contributions of faint stars and bright stars: 
$\sigma$(foreground stars)=$(1-a)\sigma$({\rm bright stars})$ + a \sigma$({\rm faint stars}), where $\sigma$ represents the surface number density.
The coefficient, $a$, 
 was determined by minimizing the fluctuation in the difference map for the Control field.
Finally we subtracted each of the foreground maps (Fig. 2S(B) and (E)) 
from each of the raw maps (Fig. 2S(A) and (D)), after scaling the
foreground map so that the mean difference for the Control field is zero.
Finally we combined the maps for the blue and red globular clusters 
to make a map for all the globular clusters.
The resulting maps are shown in Fig. S2. 
The final difference maps (Fig. 2S(C), (F), and (I)) show the spatial distribution of all, blue, and red globular clusters in Virgo, respectively.

\setcounter{figure}{0}
\renewcommand{\thefigure}{S\arabic{figure}} 

\begin{figure}
\includegraphics [width=\textwidth]{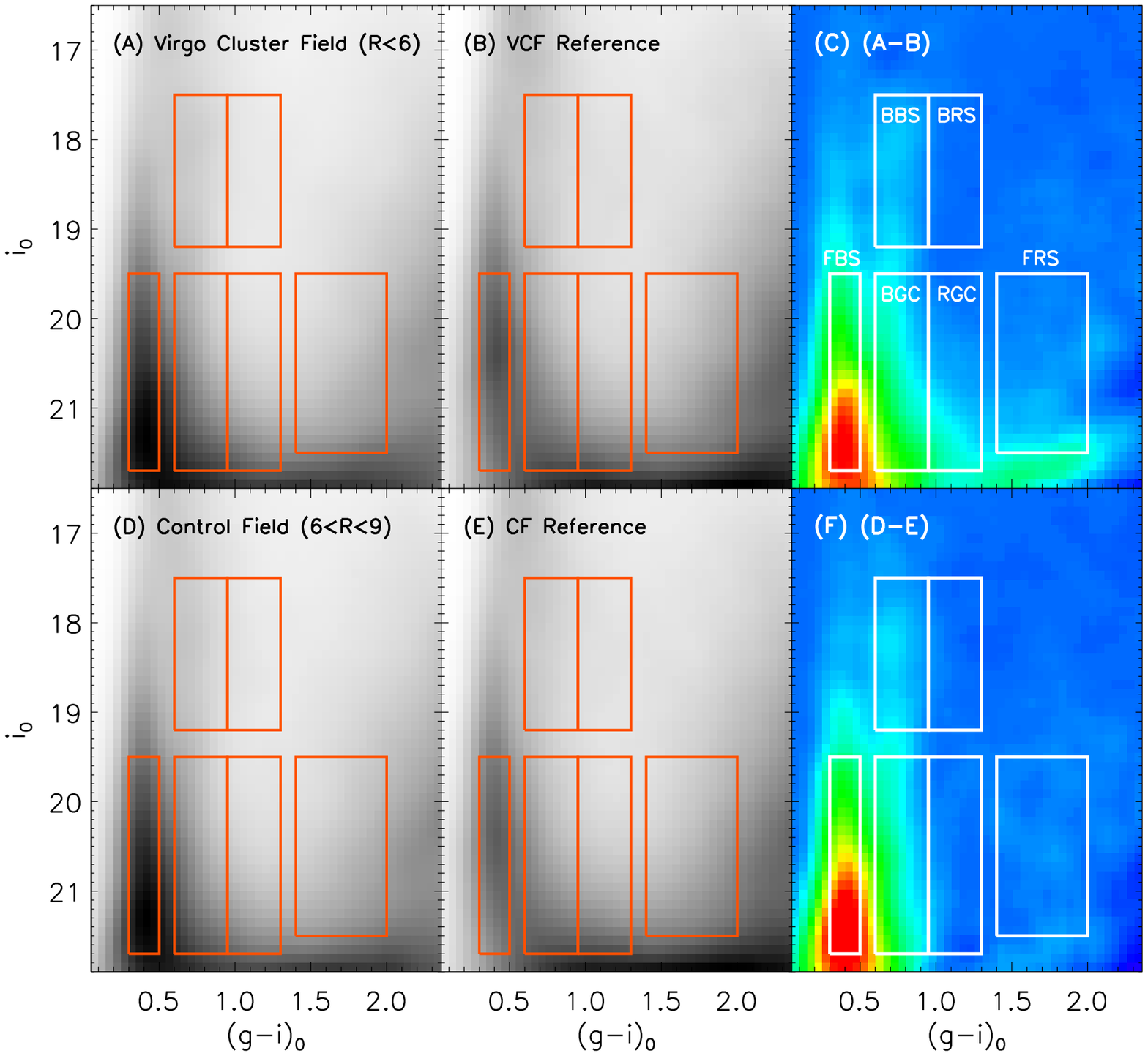} 
\caption[]{ 
$i_0 -(g-i)_0$ CMDs for 
(A) the Virgo cluster field ($R<6$ deg),
(B) the reference field with the same area as the Virgo cluster field (VCF), and
(C) the difference between (A) and (B). 
(D), (E) and (F): similarly for 
the Control field (CF) ($6<R<9$ deg). 
The boxes represent the boundaries selecting
blue globular clusters (BGC), red globular clusters (RGC), bright blue foreground stars (BBS),
faint blue foreground stars (FBS), bright red foreground stars (BRS),
and faint red foreground stars (FRS). 
Blue stars in the strong vertical structures in (C) and (F) are mostly Virgo overdensity stars. Darker or redder color represents  higher number density.
}
\label{figS1}
\end{figure}
\clearpage

\begin{figure}
\includegraphics [width=\textwidth]{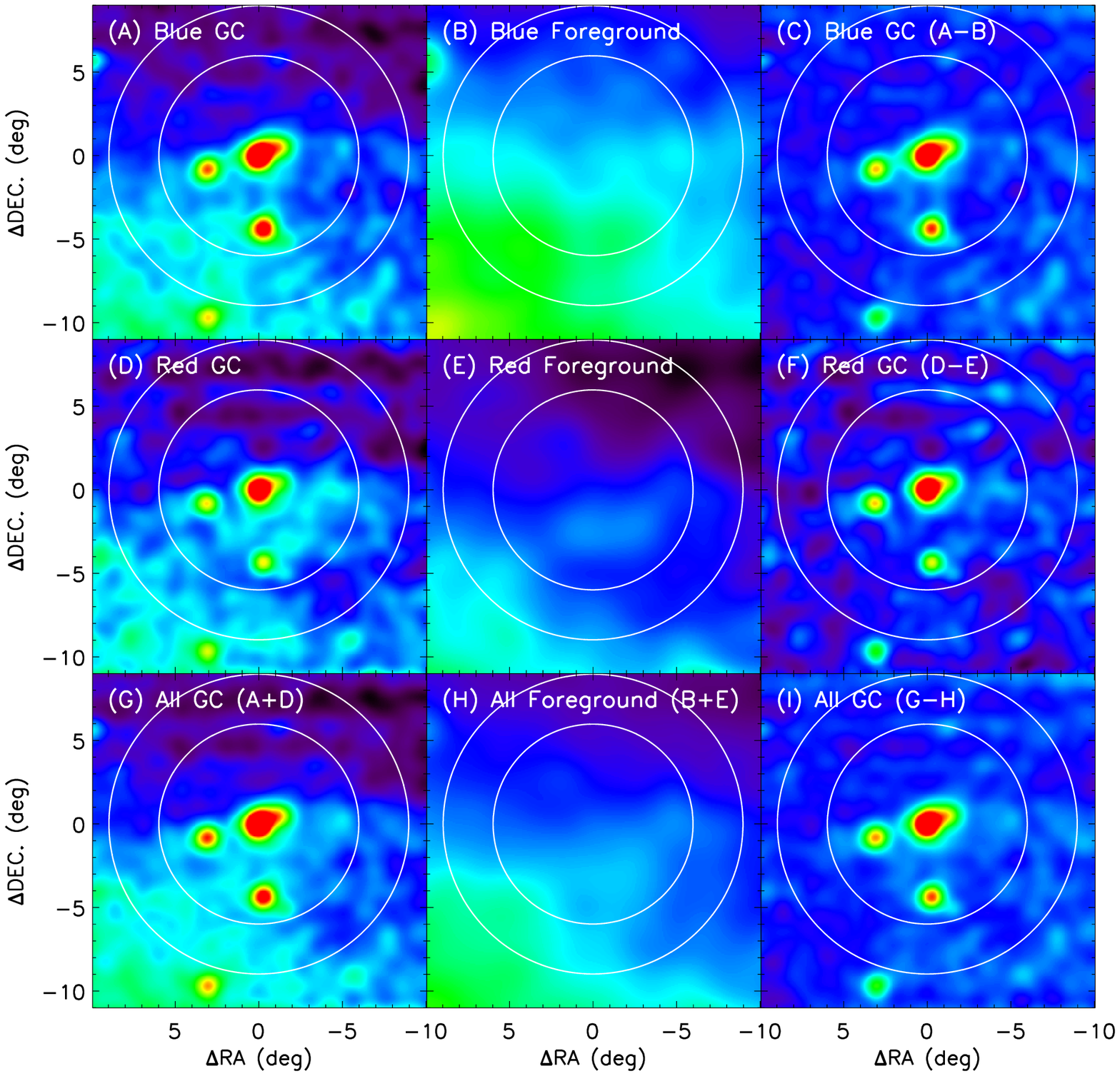} 
\caption[]{ 
Surface number density maps for (A) all the point sources with blue globular cluster color,
(B) the blue foreground stars, and (C) the difference of these two.
(D), (E), and (F): similarly for the red globular clusters.
(G), (H), and (I): similarly for all the globular clusters.
Large circles represent the boundaries for the Virgo cluster field ($R<6$ deg) and the Control field ($6<R<9$ deg). North is up and east to the left. Redder color represents  higher number density.
}
\label{figS2}
\end{figure}
\clearpage

%


\begin{thebibliography}{1}
\bibitem{note1}  
The magnitude limits for point sources in the SDSS catalog are 
22.2 mag and 21.3 mag in the $g$ and $i$-bands, respectively,  for 95 \% completeness.
\bibitem{note2} 
We derived a relation between $(g-i)_0$ color  and [Fe/H] using the SDSS photometry and [Fe/H] in the literature for the bright globular clusters M49, M60, and M87: 
[Fe/H]$=3.48(g-i)_0 - 4.13$.
\end{thebibliography}
\end{document}